\documentclass[12 pt]{article}
\usepackage{epsfig}
\usepackage{graphicx}
\begin{document}
\title{The Radiation Balance for a semi-gray Atmosphere}
\author{J.M.J. van Leeuwen}
\maketitle
\begin{center}
Instituut-Lorentz, Universiteit Leiden,\\
Niels Bohrweg 2, 2333 CA Leiden, The Netherlands.
\end{center}

\begin{abstract}
  The equations governing the radiation balance are discussed and
  exactly solved for the model of the semi-gray atmosphere, i.e.~an
  atmosphere that has regions of transparant frequencies
  and regions with the same absorption coefficient.
\end{abstract}

\section{Introduction}
The impact of human behavior on the evolution of the climate
is one of key questions of the environmental science.
In particular to explore the possibilities to divert the climate
in desired directions. There is an extensive climate science
based on big data and large scale
computations, which are hard to appreciate for outsiders
\cite{harries, zhong,he}.
So there is a gap between the professionals
and the rest of the science community. 
Thus simple models to illustrate the transport of energy
through the atmosphere are most welcome. There are several
attempts \cite{wilson}, but realistic models are soon
too complicated and simpler models, e.g. a gray atmosphere,
miss essential features.

The climate problem has two important aspects: the establisment
of a temperature gradient due to absorption of radiation and
the reaction of the atmosphere to this temperature distribution.
The radiative part is governed by a few general principles, such as
conservation of energy and local equilibrium, while the reaction
to the temperature profile induced by radiation is complex and
highly empirical. In this note we present a model for the
atmosphere, the semi-gray atmosphere, which demonstrates the
main features of the radiation balance and which is still exactly
soluble. There are good reasons to focus first on the radiative
part of the problem since radiation is fast and the temperature
on the surface of the earth follows to a large extent the
variations of the solar
input with the lattitude and with the seasonal and daily time.

For a stationary state the energy of the earth has to be
in balance: the earth
should radiate as much energy into space as it receives from
the sun. The input from the sun is given
by the geometry of the earth with respect to the sun. The
incoming radiation is in the visible  range of frequencies
for which the atmosphere is practically transparant.
On the surface of the earth the radiation is converted into heat
and that radiates back into space as thermal radiation.
The atmosphere contains greenhouse gasses, notably CO$_2$,
which absorb the radiation for the frequencies of the absorption
lines. The associate energy is re-emitted as thermal
radiation upwardly and downwardly. The stationary state
of the interplay between the  energy streams yields the radiation
balance equations. 

The simplest representation of the atmosphere is the so-called
gray atmosphere, in which all frequencies absorb equally strong
the passing radiation. This model ignores the essential feature
that the absorption is concentrated in lines separated by zones
of free transmission. The model that we treat in this note
may be characterised as a semi-gray atmosphere.
The semi-gray atmosphere has transparant regions and
regions with the same absorption coefficient.
These regions may be intermittent, mimicking the spectrum
of absorption lines. The semi-gray atmosphere has the important
mechanism of converting the radiation from absorbing
frequencies to transparant frequencies.

One may ask what is the use of only studying the radiation balance,
while important processes such as convection, evaporation and
cloud formation co-determine the magnitude of the greenhouse effect.
In particular clouds are big spoilers as they not only result from,
but also influence the radiation balance.
So we confine ourselves here to what is called a clear-sky scenario.
Fortunately one can select in the measurements the clear-sky cases.

The daily and seasonal variations in the incoming radiation,
which are directly reflected in the same variations on the surface
of the earth, demonstrate how fast and dominant the radiation is
for the climate. So it is worthwile to consider first the radiation
balance as it already gives an
important indication for the the effect of e.g.
increasing CO$_2$ on the atmosphere. The pre-industrial level of
CO$_2$ was 280 ppm and presently it exceeds 420 ppm, which is more
than a 50\% increase. The CO$_2$ concentration
has an impact on the climate only through its influence
on the radiation balance in the atmosphere. It has no direct
impact on important secondary processes. CO$_2$ owes its prominence
as greenhouse gas to the strong absorption in the thermal
radiation range. So understanding how the radiation
balance comes about is a prerequisite for answering the question
of the role of CO$_2$ on the climate.

The physics of the radiation balance
is the solvable part of the problem.
Wilson and Gea-Banacloche \cite{wilson}
provided such a simple model for treating the role
of CO$_2$. They use the temperature profile of the atmosphere as
input. In the present paper we show that for their model the
radiation
equations can be solved exactly with the temperature profile
as result. The explicit solution has many advantages as
one better sees the influence of the parameters of the model
on the radiation balance and the associate temperature profile.

Treating only the radiation part of the problem prompts the question
how to represent the effect of the increase of a greenhouse gas.
Standardly the effect is expressed as a radiative forcing.
We will show that the radiative balance implies a ratio $R$ between
the temperature at groundlevel and that of the top of the atmosphere
(TOA). $R$ depends only on the composition of the atmosphere. Since
the atmosphere is well-mixed, the same $R$ can be used everywhere
on the earth, although the variation of the temperature of the TOA
varies largely with the lattitude and the season.

We first discuss the 1-dimensional version of the model in which
the radiation is only in the vertical direction. That simplifies
the mathematics and leads to a closed expression for the
greenhouse ratio $R$. Then we consider the angular resolved model and
show how this version can be solved. Finally we construct a very
good approximation which again leads to a closed expression for $R$.
The paper closes with a discussion of the outgoing spectrum and the
significance of the result.

\section{The Model}

The semi-gray atmosphere splits the spectrum in a transparant
region and an absorbing region. For the criterion between
transparant and absorbing we consider the 
cross-section $\sigma(\nu)$ of a greenhouse molecule,
which is a function of the frequency $\nu$.
Multiplying $\sigma(\nu)$ with
the density $n(z)$ defines the inverse absorption length $l(\nu)$
\begin{equation} \label{x1}
  l(z,\nu)= \frac{1}{n(z) \sigma(\nu)}
\end{equation}
The absorption length gives the distance over which radiation
dies out. The density $n(z)$ of the atmosphere decays in an
exponential way
\begin{equation} \label{x2}
    n(z) = n(0) \exp(-z/l_{\rm atm})\equiv n(0) \tau(z),
\end{equation}
where $l_{\rm atm}$ is the ``height'' of the atmosphere (8.5 km).
$\tau(z)$ is the ``thickness'', later to be related to the
optical thickness \cite{thickness}.
The ratio $q(\nu)$ of the absorption length and $l_{\rm atm}$
\begin{equation} \label{x3}
  q(\nu) = \frac{l(0,\nu)}{l_{\rm atm}},
\end{equation} 
gives a dimensionless measure of the strength of the absorption.
We may see frequencies $q(\nu)>1$ as absorbing and $q(\nu))<1$
as transparant. The frequencies are weighted with the normalized
Planck distribution
\begin{equation} \label{x4}
  P_{\rm Pl} (\nu) = \frac{15}{\pi^4} \left( \frac{h}{k_B T} \right)^4
  \frac{\nu^3}{\exp(h \nu/k_B T) -1}.
\end{equation}
The fraction $r$ of absorbing frequencies is then given by the
integral
\begin{equation} \label{x5}
  r= \int_{\rm abs} d \nu P_{\rm Pl} (\nu),
\end{equation} 
where the integration runs over the absorbing segment.
The average absorption $\langle q \rangle$ is given by
\begin{equation} \label{x6}
  \langle q \rangle =\int d \nu q(\nu) P_{\rm Pl} (\nu).
\end{equation}
The semi-gray atmosphere is defined as transparant $q(\nu)=0$
outside the absorbing segment and absorbing $q(\nu)=q$ 
\begin{equation} \label{x7}
  q = \langle q \rangle /r
\end{equation}
inside the absorbing segment. The value of $q$ is thus chosen
such that the average absorption of the model equals that of the
full spectrum. $q$ and $r$ are the parameters of the model.
The case $r=1$ corresponds to a gray atmosphere
with a uniform absorption coefficient for the whole frequency range.

The core of the paper is the derivation of
radiation equations and the solution of these equations. With the
explicit solution we can determine the size of the greenhouse effect
as function of $q$ and $r$ as far as it is determined by the
radiation balance.
 
\begin{figure}[ht]
\begin{center}
  \epsfxsize=\linewidth  \epsffile{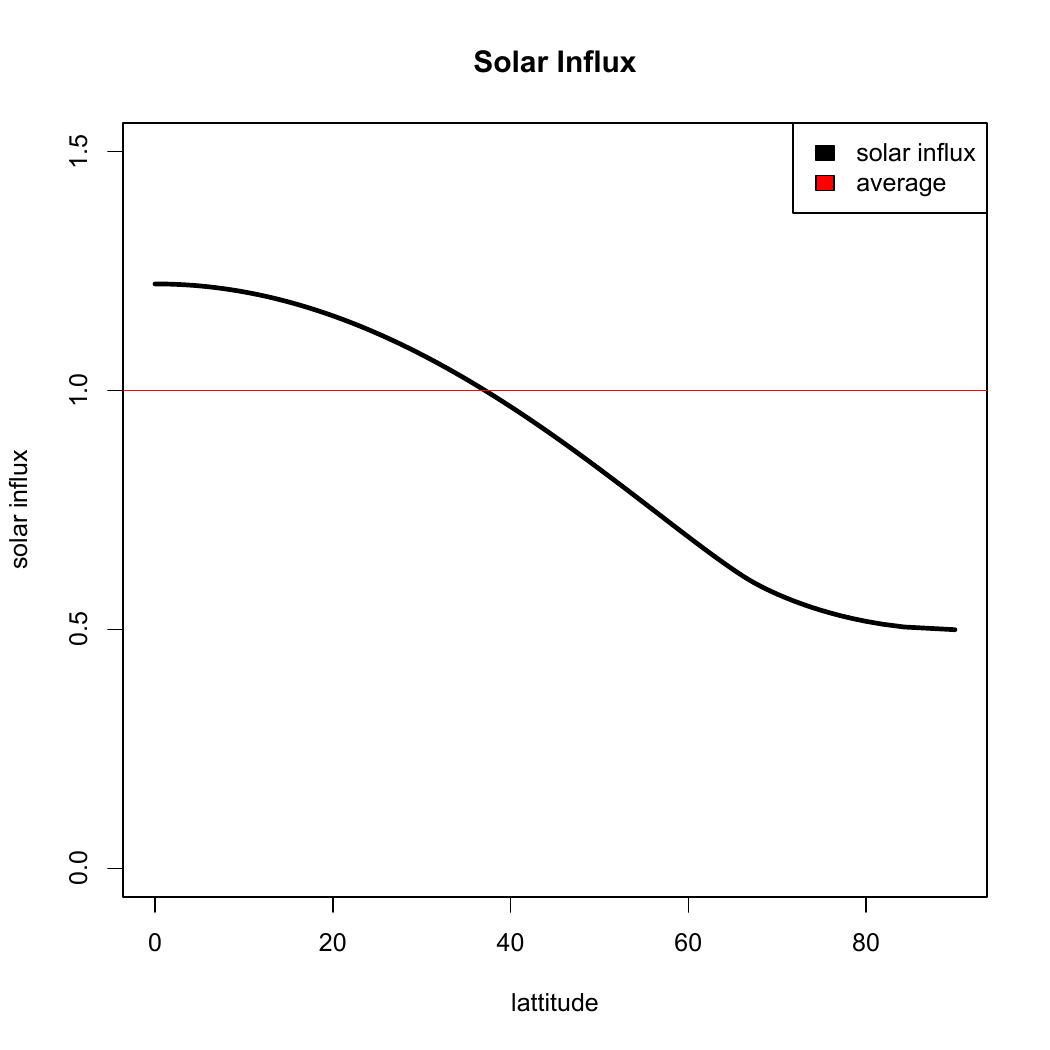}
  \caption{Yearly average of the solar radiation as function of the
    lattitude $\theta$}
\label{year}
\end{center}
\end{figure}

\section{The Radiation Balance}

The only way to get rid of the energy of the incoming solar radiation
is to radiate the same amount back into space.
There is no need to balance
the in- and outgoing radiation at every moment and everywhere. As the
radiation equations are linear it suffices to balance the radiation
on the average in space and time. One may consider the two extremes:
one where the earth is perfectly conducting such that the radiation
temperatures are the same for all positions and one where the earth is
a perfect insulator. Then the balance between and incoming and
outgoing radiation has to be satisfied for every lattitude.
The reality is in between.
The average temperatures on the various lattitudes are
quite different but not as large as the influx would give.

There are three streams of relevant thermal radiation.
The primary outgoing stream is the radiation from the
surface of the earth.
While passing through the atmosphere it is partially absorbed.
The absorbed energy is released again in the form of radiation
and generates secondary streams in upward and downward direction.
The basic equation for the strength of these streams is
the condition that at every
layer in the atmosphere the net (upward) current is constant and
equal to the incoming solar radiation.
The process of absorption an re-emission of radiation forms a
resistence for energy flow. To overcome this resistence a
radiation pressure must build up by a gradual temperature increase
towards the ground. From the equations for the radiation flow
we derive the temperature distribution of the atmosphere.
The difference between the ground temperature
and that at the top of the atmosphere (TOA) gives the radiation
greenhouse effect. Due to other parallel energy flows (convection,
turbulence, latent heat of evaporation) the true greenhouse
effect is smaller than the radiation greenhouse effect.

Before we discuss the model equations
we first collect some data on the input
of the solar radiation. Then we introduce the radiation equations
for a one-dimensional vertical flow (see also \cite{wilson}).
This simplified model better
shows the competition between the various flow components.
Moreover the solution of these balance equations can be obtained
by pure analytical means. Next we discuss the three-dimensional
flow pattern.  Finally we analyze how the
density of CO$_2$ influences the parameters $q$ and $r$ and draw
the conclusion.

\section{The radiation input}

The basic relation between temperature and black body radiation is the
law of Stefan-Boltzmann
\begin{equation} \label{a1}
  F=\sigma_{\rm SB} T^4
\end{equation}
Here $T$ is the absolute temperature of the radiating substance, $F$
the outgoing radiation flux and $\sigma_{\rm SB}$ is a universal
constant with the value
\begin{equation} \label{a2}
  \sigma_{\rm SB}  = 5.670374419 * 10^{-8} {\rm W}/{\rm m}^2
  {\rm K}^{-4}.
\end{equation} 
$\sigma_{\rm SB}$ involves only fundamental constants like
the speed of light, the Planck constant and the Boltzmann constant.

The input radiation is determined by the surface temperature of the
sun $T_\odot =5778$ K and a number of geometrical factors.
The flux of solar radiation $S_\odot$ is associated with the
temperature of the sun 
\begin{equation} \label{a3}
  S_\odot = \sigma_{\rm SB} T^4_\odot = 5.6704*10^{-8} (5778)^4 =
  6.319*10^7 \, {\rm W}/{\rm m}^2.
\end{equation}  
The radius of the sun $R_\odot=6.957*10^8$ m and the distance of
the earth to the sun is $R_\oplus =1.496*10^{11}$ m. Thus the
strength of the flux at the earth is diminuished to
\begin{equation} \label{a4}
S_\oplus = S_\odot (R_\odot /R_\oplus)^2 = 1.369*10^3 \, {\rm W/m}^2.
\end{equation}
The earth has a cross-section for the radiation of $ \pi R^2_O$ and
a surface $4 \pi R^2_O$. So on the average the earth receives the flux
$S_\oplus/4$. Moreover the earth has an albedo which fluctuates
remarkably stable around 0.29 \cite{albedo}.
This means the 29\% of the incoming flux is reflected and
71\% reaches the surface of the earth. So the effective in-flux is
\begin{equation} \label{a5}
S_{\rm in} = 0.71 S_\oplus /4 = 243 \,   {\rm W/m}^2 \equiv F_{\rm out},
\end{equation}
which heats the earth. The problem is how it is transmitted
through the atmosphere back into space.

A simplistic view, due to Arrhenius \cite{arrhenius},
is to see the earth as a
black body radiator with a uniform temperature $T_{\rm out}$.
According to the Stefan-Boltzmann law $T_{\rm out}$ should be
\begin{equation} \label{a6}
T_{\rm out} = (F_{\rm out}/\sigma_{\rm SB})^{1/4} = 256 \, {\rm K}.
\end{equation} 
Compared to the average ground temperature of 288 K this differs 32 K,
which is attributed to the greenhouse effect. 
In Fig.~\ref{year} we plot the incoming radiation as function
of the lattitude.

The differences are large. At the equator the influx is 25\% larger
than average and at the poles 50\% smaller than average. 
The greenhouse effect is the temperature difference between
the surface
of the earth and the top of the atmosphere where the air is so dilute
that thermal radiation passes without noticable absorption.
There the outgoing flux must equal the flux given by Eq.~(\ref{a5})
in order that the earth keeps the same temperature.

An alternative and
more useful measure for the greenhouse effect is the ratio $R$ of the
temperature at the surface of the earth and the temperature at the
top. This is a property of the composition of the atmosphere. The
estimate of Arrhenius gives for the present value $R$=288/256=1.125.
The fourth power $R^4$ gives the ratio of the intensity of the
radiation at the surface and the top. Since the equation for the
radiation is linear we may use $R$ to estimate the ground
temperature for different values of the top radiation.
\begin{figure}[ht]
\begin{center}
  \epsfxsize=\linewidth  \epsffile{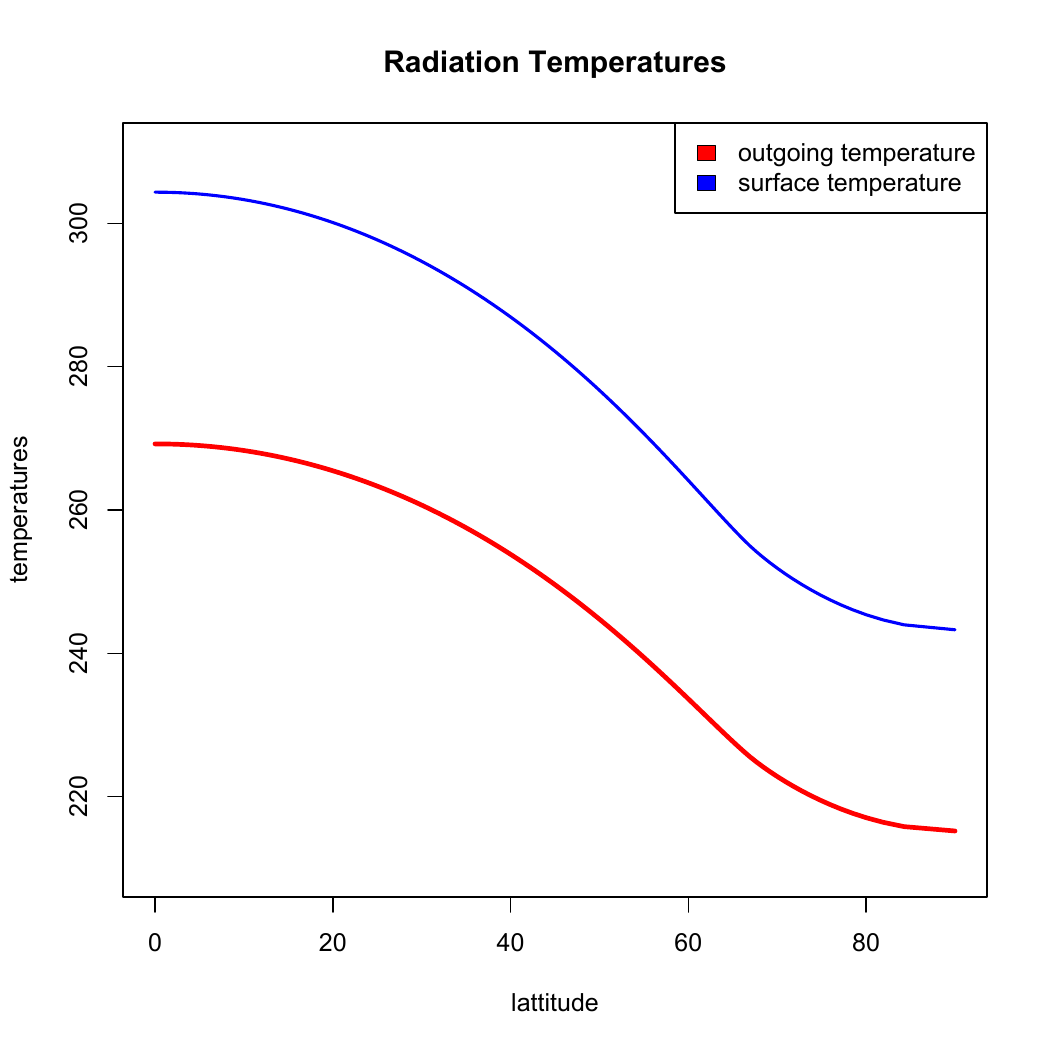}
  \caption{Yearly average of the radiation as function of the
    lattitude $\theta$}
\label{Temps}
\end{center}
\end{figure}

An example of such different values of the outgoing radiation
follows from the opposite of the Arrhenius assumption.
Arrhenius assumes that the
temperature on the earth is everywhere the same as it would be if the
earth would be a perfect conductor. The opposite is to assume that
the earth is a perfect insulator. Then there is no exchange of heat
between the various lattitudes, which receive quite different amounts
of radiation.
Let $\phi (\theta)$
be the angle between the direction of the sun and the normal of the
surface at a lattitude $\theta$. We use the function
$f(\theta)$ as
\begin{equation} \label{a7}
  f(\theta) = 4 \langle \cos(\phi (\theta)) \rangle, 
\end{equation}
where the average is over the duration of the day and the days of the
year. The 4 is for convenience since we know that the average over all
lattitudes equals 1/4. Thus $f(\theta)$ gives the fraction of the
radiation that enters at lattitude $\theta$.
Then we have the thermal out-flux $F_{\rm out}(\theta) $ equal to
solar influx at lattitude $\theta$
\begin{equation} \label{a8}
  F_{\rm out} (\theta) = S_{\rm in} \, f(\theta).
\end{equation}
with $S_{\rm in}$ given by Eq.~(\ref{a5}). The same amount has to
leave at the top of the atmosphere. The atmosphere has at all
lattitudes the same composition. So we may use the same ratio $R$
for all lattitudes. Thus the surface temperature at lattitude $\theta$
equals
\begin{equation} \label{a9}
  T(\theta) = R \, T_{\rm out} \, f(\theta)^{1/4}= 288\,
  f(\theta)^{1/4}  {\rm K}.
\end{equation}

The curve $f(\theta)^{1/4}$ is also shown in Fig.~\ref{Temps}. Although
the variation in $f(\theta)$ is substantial, the 1/4 power makes the
deviations from 1 much smaller. Translating them with Eq.~(\ref{a9})
into temperatures we find 31 $^o$C at the equator and -30 $^o$C at the
poles. In reality the differences are smaller as can be expected since
heat flows from the equator to the poles leveling the temperature
differences. But this temperature profile is closer to the reality
than the constant value of Arrhenius.

\section{The absorption Equation}

The surface of the earth emits thermal radiation in response to the
influx of solar radiation in the visible frequency range. In this
section, where we consider only radiation in the vertical direction,
the radiation components are functions of the height $z$
and the frequency $\nu$. The primary stream is the radiation of the
surface of the earth.

\subsection{The Primary stream}

The primary stream  is denoted as $I_0(z; \nu) $.
At the groundlevel $z=0$ it has a frequency distribution 
given by the (normalized) Planck distribution (\ref{x1}).
Thus at $z=0$ $I_0(0; \nu) $ has the shape
\begin{equation} \label{b1}
  I_0(0; \nu) = I_0 \, P_{\rm Pl} (\nu)
\end{equation}
On its way to higher layers it is partially absorbed by the greenhouse
gasses with absorption cross-section $\sigma(\nu)$.
The absorption causes the radiation to decay in an exponential way,
the law of Lambert-Beer \cite{lambert}
\begin{equation} \label{b2}
  I_0( z; \nu) = G ( z,0; \nu)  I_0(0; \nu),
\end{equation}
where the propagator $G$ is given by
\begin{equation} \label{b3}
  G ( z,z'; \nu) =\exp \left(- \int^z_{z'} 
    d z'' n(z'') \sigma (\nu)\right).
\end{equation}
The integration in the exponent of $G$ can be carried out
\begin{equation} \label{b4}
  G ( z,z'; \nu) = \exp \left(-q(\nu)|\tau(z)- \tau(z')| \right).
\end{equation}
In our model $q(\nu)$ is replaced by 0 for the transparant
region and by $q$ for the absorbing region.

\subsection{The Secondary Streams}

The greenhouse molecules, like CO$_2$, collide at  a high rate
($10^7$/sec) with the ambient air molecules. These collisions 
broaden the absorption lines to Lorentzian lines \cite{lorentz}.
The absorption is proportional to the strength $I(z; \nu) $
of the radiation at level $z$ multiplied by the 
absorption coefficient $n(z) \sigma(\nu)$. The excited greenhouse
molecules collide again frequently with the air molecules before
they would fall back to the groundstate. So the excitation energy
leaks away to the degrees of freedom of all the other molecules
and the extra energy is radiated away as thermal energy,
distributed in frequency by a Planck distribution.
The emitted radiation
has an amplitude\footnote{The dimension of $J$ is flux/length}
$J(z)$ and is thermally distributed over the frequencies.
Conservation of energy then requires that
\begin{equation} \label{c1}
  J(z) dz = n(z) dz  \int d \nu \sigma(\nu) \,  I(z; \nu).
\end{equation}
On the right hand side we have the amount of energy absorbed
in the layer ($z,dz$) and on the left hand side the amount
of energy that is emitted in layer ($z,dz$) \cite{kirchhoff}.

The incoming flux $I(z; \nu) $ is the sum of three currents
\begin{equation} \label{c2}
  I( z; \nu) = I_0 ( z; \nu) +
  I_\uparrow (z; \nu) + I_\downarrow (z;\nu).
\end{equation}
The first is the primary current coming from the surface of the
earth defined in Eq.~(\ref{b4}). $I_\uparrow$ is the secondary upstream
coming from the layers below $z$ and $I_\downarrow$ is the downstream
coming from the layers above $z$.

The upstream $I_\uparrow$ is the integral of the emitted radiation at
level $z'<z$ in the upward direction
\begin{equation} \label{c3}
  I_\uparrow (z; \nu) = \frac{1}{2} \int^z_0 dz' 
  G( z,z'; \nu) J(z') P_{\rm Pl} (\nu).
\end{equation}
The strength of the emitted radiation at level $z'$ equals $J(z')$
and it has a thermal distribution over the frequencies $\nu$.
Likewise the downstream
\begin{equation} \label{c4}
  I_\downarrow (z; \nu) = \frac{1}{2} \int^\infty_z dz'
  G(z,z';\nu ) J(z')P_{\rm Pl} (\nu).
\end{equation}
$J$ is expressed by Eq.~(\ref{c1}) in terms of $I(z; \nu)$,
which is the sum of the three currents $I_0, I_\uparrow$ and
$I_\downarrow$, which in turn depend by Eq.~(\ref{c3})
and (\ref{c4}) on $J$. So Eq.~(\ref{c1}) is an integral equation for
$J$. 
The study of this integral relation is the core of this note. We will
refer to it as to the {\it absorption} equation since it expresses the
equality between the absorbed and re-emitted radiation.

For the semi-gray atmosphere the integration over the frequencies
can be carried out in the absorption equation (\ref{c1}).
The factor $\sigma(\nu)$ selects the absorbing domain $r$. Thus
\begin{equation} \label{c5}
  J(z) = \frac{q r\tau(z)}{l_{\rm atm}} \,
  \left[I_0^a(z) +I_\uparrow^a (z) +I_\downarrow^a(z) \right],
\end{equation}
where the superscript $a$ indicates that the absorbing component
of the current has to be taken \cite{kirchhoff}
The three integrated currents read
\begin{equation} \label{c6}
  \left\{
    \begin{array}{rcl}
      I_0^a(z) & = & I_0 G(z,0), \\*[4mm]
      I_\uparrow^a (z)&  = & \displaystyle\frac{1}{2} \int^z_0 dz'
            G(z,z') J(z'), \\*[4mm]
      I_\downarrow^a (z) & = &\displaystyle \frac{1}{2} \int^\infty_z
             dz' G( z,z') J(z').
    \end{array}
  \right.
\end{equation}
The propagator $G(z,z')$ is given in Eq.~(\ref{b4}) with the $q$
of the model.
\begin{equation} \label{c7}
  G ( z,z') = \exp \left(-q|\tau(z)- \tau(z')|  \right).
\end{equation}
The equation becomes more transparant by changing from the height
variable $z$ to the optical thickness $\rho(z)$  
\begin{equation} \label{c8}
  \rho(z)=q \tau(z) = q \exp(-z/l_{\rm atm}), \quad \quad \quad
  z(\rho)=-l_{\rm atm}
  \log(\rho/q), \quad \quad \quad dz=-l_{\rm atm} \frac{d \rho}{\rho},
\end{equation}
which leads to the relation
\begin{equation} \label{c9}
  j(\rho) = I_0 r \exp(\rho-q) + \frac{r}{2} \int^q_\rho d \rho'
  \exp(\rho-\rho') \, j(\rho') +\frac{r}{2} \int^\rho_0 d \rho'
  \exp(\rho'-\rho) \, j(\rho'),
\end{equation}
where $j(\rho)$ stands for the combination
\begin{equation} \label{c10}
  j(\rho)=l_{\rm atm} J(z(\rho)) /\rho.
\end{equation}
Eq.~(\ref{c10}) can be solved analytically. Before we give the
solution we compare it with another equation expressing the
conservation of energy.

\section{The Flow Equation}

Since there is no accumulation or loss of energy in the layers of the
atmosphere the net upward flow of radiation is constant
and equal to the outward flow of energy at the top of the atmosphere
\cite{schwarzschild}
\begin{equation} \label{d1}
  F(z) = I_0 (z) +I_\uparrow(z) - I_\downarrow(z)=F_{\rm out}.
\end{equation}
On the average over the earth it equals $S_{\rm in}$ given in
Eq.~(\ref{a5}). As the ingredients of the flux $F(z)$ are
linearly dependent on $J(z)$, Eq.~(\ref{d1}) is also an
integral equation for $J(z)$. We call this the {\it flow} equation.
In order to see the relation between the absorption equation and
the flow equation we make the contributions to Eq.~(\ref{d1})
explicit for the semi-gray model. Each of the three streams has
an absorbing and transmitting part
\begin{equation} \label{d2}
  I_\gamma(z) = r I_\gamma^a(z) + (1-r) I_\gamma^o(z), \quad \quad
  \gamma = 0, \uparrow, \downarrow.
\end{equation}
The absorbing components are given in Eq.~(\ref{c6}) and the
the transmitting components are given by the same expression
Eq.~(\ref{c6}) but with $G(z,z')$ replaced by 1. Using the
transformation Eq.~(\ref{c8}) we find for the three streams
\begin{equation} \label{d3}
  \left\{
    \begin{array}{rcl}
      I_0(z) & = & I_0 [r \exp(\rho -q) + (1-r) ], \\*[4mm]
      I_\uparrow(z) & = & \displaystyle \frac{r}{2} \int^q_\rho d \rho'
             \exp(\rho-\rho') \, j(\rho') + \frac{1-r}{2}
             \int^q_\rho d \rho' j(\rho'), \\*[4mm]
       I_\downarrow(z) & = & \displaystyle \frac{r}{2} \int^\rho_0 d \rho'
             \exp(\rho'-\rho) \, j(\rho') + \frac{1-r}{2}
                             \int^\rho_0 d \rho' j(\rho').
    \end{array}
  \right.
\end{equation} 
Using the expressions (\ref{d3}) for the currents in (\ref{d1})
we find that the relation
\begin{equation} \label{d4}
  \frac{\partial F(z)}{\partial \rho} = 0
\end{equation}
is the same as the absorption equation (\ref{c9}).

\section{The Solution of the Equations}

In the Appendix the solution for the absorption and flow
equation has been worked out for a slightly more general kernel
for the integral equation. Here we summarize the
solution for the parameters $a=b=1$
in Eq.~(\ref{A1}). As the absorption equation is implied by
the flow equation it suffices to solve the latter.
However the absorption equation is more suggestive
for the form of the solution which reads
\begin{equation} \label{e1}
  j(\rho) = [a_+ \exp(\alpha \rho) + a_-\exp(-\alpha \rho)]
  F_{\rm out}. 
\end{equation}
The exponent $\alpha$ equals
\begin{equation} \label{e2}
  \alpha= \sqrt {1-r}.
\end{equation}
Inserting this  solution into the absorption equation
gives two linear relations between the coefficients $a_\pm$ and
the amplitude $I_0$. Inserting the form (\ref{e1}) into the
flow equation leads to the same two relations and in addition to
a third relation in which also $F_{\rm out}$ features. Solving
the set of three linear relations between $a_\pm$ and $I_0$
gives the solution
\begin{equation} \label{e3}
  a_+ = \frac{2(1+\alpha)}{\alpha \det(q,\alpha)},
  \quad\quad
  a_- = -\frac{2(1-\alpha)}{\alpha \det(q,\alpha)}.
\end{equation}
with the expression for $\det(q,\alpha)$
\begin{equation} \label{e4}
  \det(q,\alpha) = \frac{1+\alpha}{1-\alpha} \exp(q \alpha) +
  \frac{1-\alpha}{1+\alpha} \exp(-q \alpha) +2.
\end{equation}

With the explicit form (\ref{e1}) for $j(\rho)$ all streams
can be calculated.

\section{Local Thermodynamic Equilibrium}

Sofar the discussion concerns only the radiation, but for the
greenhouse effect the temperature is important. The connection
between radiation and temperature is established through the
requirement of local thermodynamic equilibrium. In our case
it means that the total strength of the radiation $I (z)$\footnote{Rather than introducing
  new symbols for integrated quantities,
  we adopt the convention to use the same
  symbol but give it less arguments. So $I(z)$ is
  $I(z;\nu)$  integrated over the frequencies.}
\begin{equation} \label{f1}
  I(z) = I_0 (z) +I_\uparrow(z) + I_\downarrow(z).
\end{equation}
corresponds to the local temperature $T(z)$ according to the law
of Stefan-Boltzmann
\begin{equation} \label{f2}
  I(z) = \sigma_{\rm SB} T^4(z).
\end{equation}
Since $I_\downarrow(\infty)=0$ the value of
$I(\infty)=F(\infty)= F_{\rm out}$.

The greenhouse effect is best represented by the ratio of the
temperature at the surface and that at of the top of the atmosphere.
\begin{equation} \label{f3}
  R=\frac{T(0)}{T(\infty)} = \left(\frac{I(0)}{I(\infty)}
  \right)^{1/4}.
\end{equation}
Combining the radiation intensity $I(z)$ with the flow equation
(\ref{d1}) we find the useful relation
\begin{equation} \label{f4}
  I(z) = F_{\rm out} +2  I_\downarrow(z)
\end{equation}
Since $I_\downarrow(\infty)=0$ we can rewrite Eq.~(\ref{f3}) as
\begin{equation} \label{f5}
  R = \left(1+ 2\frac{I_\downarrow(0)}{F_{\rm out}} \right)^{1/4}.
\end{equation}
For the down flow $I_\downarrow(z)$ we find (using relation
(\ref{A7}) for some simplification)
\begin{equation} \label{f6}
  2\frac{ I_\downarrow(z)}{F_{\rm out}} = (\exp (\alpha \rho) - 1) a_+
    +(\exp (-\alpha \rho) - 1) a_-
\end{equation}

\begin{figure}[ht]
\begin{center}
  \epsfxsize=15cm  \epsffile{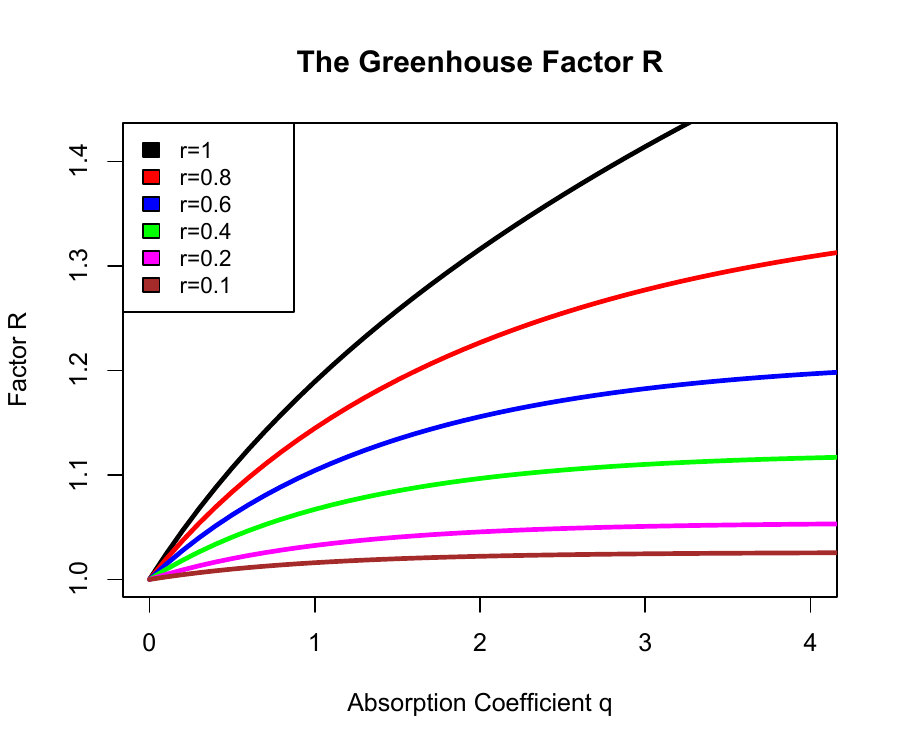}
  \caption{The Greenhouse Factor $R(q,r)$ as function of the
    absorption coefficient $q$ (optical thickness) and the absorption
    range $r$.}
  \label{trans}
\end{center}
\end{figure}

Inserting the expressions (\ref{e3}) for $a_\pm$ yields for the
ratio
\begin{equation} \label{f7}
  \frac{I_\downarrow(0)}{F_{\rm out}} = \left(\frac{1+\alpha}{\alpha}
    \exp(q \alpha)-\frac{1-\alpha}{\alpha}
    \exp(-q \alpha) -2\right) /\det(q,\alpha).
\end{equation}
In the limit of small $q$  we find
\begin{equation} \label{f8}
  R_{q \rightarrow 0} = (1 + q r)^{1/4}
\end{equation}
and for large $q$
\begin{equation} \label{f9}
  R_{q \rightarrow \infty} =\left(\frac{1-\alpha}{\alpha}\right)^{1/4}.
\end{equation} 
In Fig.~\ref{trans} we show the greenhouse factor $R$ as function
of model parameters $q$ and $r$. 
  
\section{The angular dependence of the Radiation}

Sofar we treated the atmosphere as 1-dimensional with only a vertical
coordinate. On the average the radiation propagates indeed only
in the vertical direction. But the propagation of
the average radiation is not equal to the average of the propagated
radiation, since the propagator depends on the angle of the radiation.
So we have to consider the radiation as a function, not only of height
and frequency, but also of the angle $\theta$ with the normal. The
most important change is that we have to substitute for
the propagator in the direction $\theta$
\begin{equation} \label{g1}
  G(z,z') = \exp(-|\rho(z) - \rho(z')|) \Rightarrow
  \exp(-|\rho(z) - \rho(z')|/\cos(\theta)).
\end{equation}
The $\cos(\theta)$ in the denominator of the exponent accounts for
the longer path from layer $z'$ to $z$ in the direction $\theta$.

In the absorption equation we need to integrate over all directions
$\theta$ in order to find the total energy that is absorbed.
So we encounter integrals of the type
\begin{equation} \label{g2}
  \int^{\pi/2}_0 \sin(\theta) d \theta \cos^n (\theta)
  \exp(-|\rho(z) - \rho(z')|/\cos(\theta))= E_{2+n}(|\rho(z) - \rho(z')|),
\end{equation}
where the $E_n$ are logaritmic integrals defined as
\begin{equation} \label{g3}
  E_n (x) = \int^\infty_1  \frac{dy}{y^n} \exp (- x y).
\end{equation}

The absorption equation (\ref{c9}) involves an unweighted integral
over angles ($n=0$) and therefore reads
\begin{equation} \label{g4}
  j(\rho) = I_0 r E_2(q-\rho) + r \int^q_\rho d \rho'
  E_2(\rho'-\rho) \, j(\rho') + r \int^\rho_0 d \rho'
  E_2(\rho-\rho') \, j(\rho').
\end{equation}

For the flow equation the angular integral involves an extra
$\cos(\theta)$ since only the flow in the vertical direction
counts in the upward flow. So we get the function $E_3$ for the
absorbing terms and a factor $1/2$ for the non-absorbing terms 
The three currents read
\begin{equation} \label{g5}
  \left\{
    \begin{array}{rcl}
      I_0(z) & = & \displaystyle I_0 [r E_3(q-\rho) + \frac{1-r}{2} ], \\*[4mm]
      I_\uparrow(z) & = & \displaystyle r \int^q_\rho d \rho'
             E_3(\rho'-\rho) \, j(\rho') +\frac{1-r}{2}
             \int^q_\rho d \rho' j(\rho'), \\*[4mm]
       I_\downarrow(z) & = & \displaystyle r \int^\rho_0 d \rho'
             E_3(\rho-\rho') \, j(\rho') + \frac{1-r}{2}
                             \int^\rho_0 d \rho' j(\rho').
    \end{array}
  \right.
\end{equation}
With these three currents the flow is constructed as in
Eq.~(\ref{d1}).

As a check on the equations we show that the absorption equation
is the derivative of the flow equation with respect to $\rho$.
The derivative gives two types of terms: those originating from
$\rho$ appearing in the boundaries of the integrations and those
coming from the derivative of $E_3(\rho-\rho')$. The former yield
$-j(\rho)$, which is the left hand side of Eq.~(\ref{g4}) and the
latter are computed with
\begin{equation} \label{g6}
  \frac{d E_3(\rho-\rho')}{d \rho} = - E_2(\rho-\rho')
\end{equation}

\begin{figure}[ht]
\begin{center}
  \epsfxsize=16cm  \epsffile{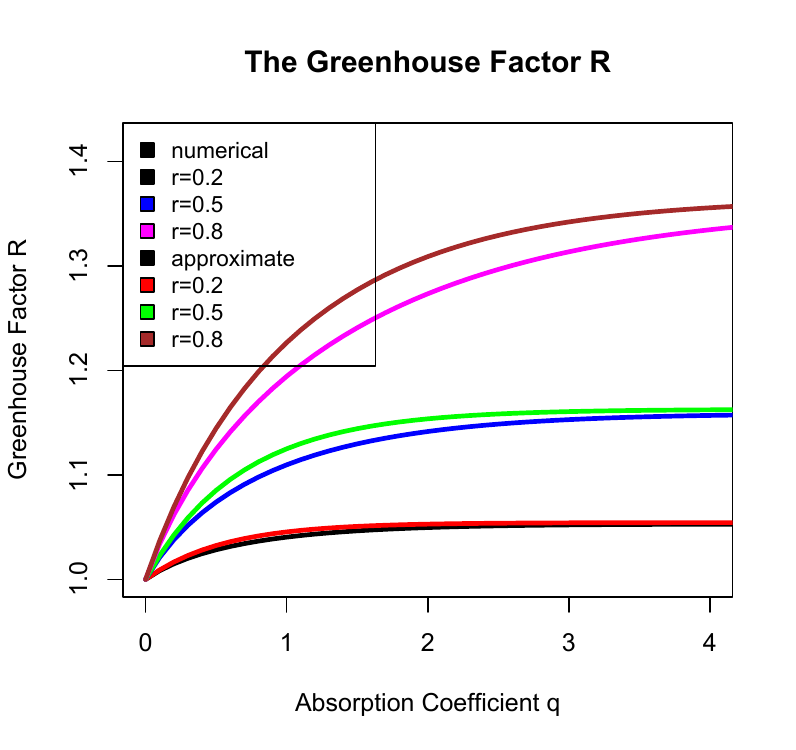}
  \caption{The Greenhouse Factor $R(q,r)$ as function of the
    absorption coefficient $q$ (optical thickness) and the absorption
    range $r$, for the angular resolved equations.}
  \label{gangular}
\end{center}
\end{figure}

We have not found a solution for Eq.~(\ref{g4}) in closed form.
Writing the integrals as sums over a set of points, the integral
equation becomes a set of linear equations which permit a mumerical
solution. In Fig.~\ref{gangular} we show the greenhouse factor $R$
for the angular resolved equations.
In the same Figure we also have drawn the results for the approximation
\begin{equation} \label{g7}
  E_3(\rho-\rho') \Rightarrow \frac{1}{2}\exp(-2(\rho-\rho')), \quad
  \quad
  E_2(\rho-\rho') \Rightarrow \exp(-2(\rho-\rho'))
\end{equation}
These replacements are correct for $\rho=\rho'$. For $E_2$ the
integral is the same and for $E_3$ the integral, which is 1/3,
becomes in the appoximation equal to 1/4. The equations can be
solved in closed form for this approximation (see the Appendix).
One sees in the Fig.~\ref{gangular} that the approximation is 
very accurate for low $r$. For $r=0.8$ the approximation still
follows the trend but is less accurate. The approximate solution
just has $q$ replaced by $2q$ in the 1-dimensional model.

\section{The Spectrum of the outgoing Radiation}

Another signature of the greenhouse effect is the change in the
spectrum of the outgoing radiation. Without absorption the spectrum
would be distributed according to the Planck distribution, with the
temperature based on the strength of $F_{\rm out}$. Due to the
absorption the absorbing frequences are depleted in favor of the
transparant frequencies. The depleted frequences are represented by
\begin{equation} \label{i1}
  I_0^a (\infty) + I_\uparrow^a (\infty) = \frac{4}{\det(q,\alpha)}
  F_{\rm out}
\end{equation}
The transparant frequencies that benefit of the absorption are given
by
\begin{equation} \label{i2}
  I_0^o (\infty) + I_\uparrow^o (\infty) =\left(1 -
    \frac{4}{\det(q,\alpha)}\right) F_{\rm out}
\end{equation}
Note that the sum of the currents on the left hand side of
Eq.~(\ref{i1}) and (\ref{i2}) equals $F_{\rm out}$ as it should
on the basis of the flow equation. Since
$\det(q, \alpha)$ grows with the absorption $q$ the part of
the absorbing spectrum gradually decreases in favor of the
transparant part.

\begin{figure}[ht]
\begin{center}
  \epsfxsize=15cm  \epsffile{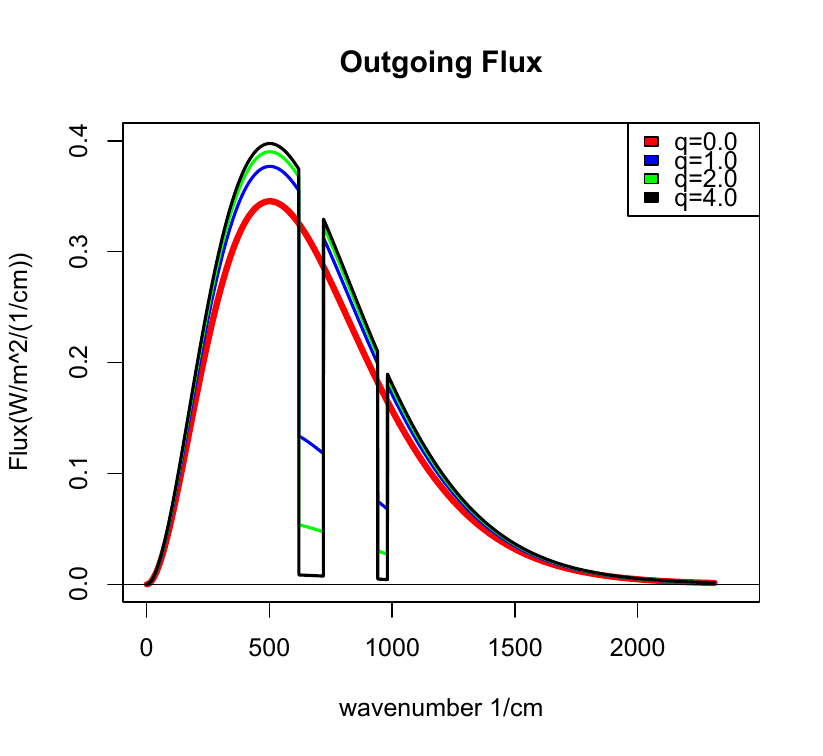}
  \caption{The spectrum of the outgoing radiation for a number of
  absorptions $q$ and a value of $r=0.14$ }
  \label{spectrum}
\end{center}
\end{figure}

We get the intensities of the radiation in the two regimes
by dividing them by their size: $r=1-\alpha^2$ for the absorbing
part and $1-r= \alpha^2$ for the transparant part. So the spectrum
reads for the transparant part
\begin{equation} \label{i 3}
  I^o(\nu; \infty) = \frac{1}{\alpha^2}\left(1 -
    \frac{4}{\det(q,\alpha)}\right) P_{\rm Pl} (\nu),
\end{equation}
and for the absorbing part
\begin{equation} \label{i4}
  I^a(\nu; \infty) = \frac{4}{(1-\alpha^2)
    \det(q,\alpha)} P_{\rm Pl} (\nu),
\end{equation}
In Fig.~\ref{spectrum} we have shown the outgoing spectrum for a
number of absorptions $q$ and a value of $r=0.11$.
The radiation of the
absorbing frequencies is reduced in favor of the radiation of
the transparant frequencies. This is clearly demonstrated in
Fig.~\ref{spectrum} where we show the effect of an absorbing
region between the frequencies 620 cm$^{-1}$ and 720 cm$^{-1}$
and the frequencies between 1941 cm$^{-1}$ and 1981 cm$^{-1}$.
These regions corrrespond to the absorption band of symmetric
bending mode of CO$_2$ and the transition from the first
excited symmetric stretching mode to the
first excited asymmetric stretching mode \cite{witteman}.
Again one observes that the effect of
more CO$_2$ saturates above $q=4$, which is still a rather weak
absorption with an absorption length of some 3 km.

The dent in the spectrum has sharp walls since we have a sharp
distinction between absorbing and transparant frequencies. One
observes that for $q>4$ the absorbing frequencies are almost
completely
removed from the spectrum (saturation), while $q=4$ is still a
weak absorption with an absorption length of 3 km.

\section{Conclusion} \label{conclusion}

We have studied the radiative balance of a semi-gray atmosphere,
consisting out of a region with a constant absorption coefficient
and a background of transparant frequencies. The absorbing region
may exist out of a number of separately absorbing lines. For
the exact solution it is necessary that the absorbing lines
are of equal strength. Such an atmosphere is characterized by
two parameters: the strength $q$ of the absorption and the size
$r$ of the absorbing region.

The greenhouse effect is represented as a ratio $R$
of the temperature at the surface of the earth and at the top
of the atmosphere, where the temperature is determined by the
radiative outflow. $R$ is a more convenient parameter than the
usual temperature difference between the bottom and the top of the
atmosphere, which is e.g. lattitude dependent, while $R$ only
depends on the composition of the atmosphere. 
The dependence of $R$ on $r$ and $q$ is shown in Fig.~\ref{trans}.
The general tendency is that for fixed and small $r$ the ratio $R$
increases with $q$ but soon saturates. For $r=1$
(gray atmosphere) it keeps rising with $q$.

The spectrum of the outgoing radiation gives a  more detailed
picture of the effect of the absorption. A higher
concentration of CO$_2$  has also influence on the width of the
absorbing region. According to Arrhenius \cite{arrhenius}
the width increases logarithmically with the concentration.

We have confined ourselves to the result of the radiation balance
and ignored all secondary processes, such as convection, turbulence
and the latent heat of evaporation of water. All these processes
lead to contributions to the upward energy stream. That means that
the energy current due to radiation becomes smaller for a balance
and the associated temperature gradient becomes smaller. Thus the
greenhouse effect on the basis of the radiation
balance overestimates the effect. Also there may be an unbalance due
to energy storage in the oceans. This will also lower the real
greenhouse effect.

{\bf Acknowlegdement} The author is indebted to Bernard Nienhuis for
numerous discusssions on radiation and to Henk Lekkerkerker for
arousing interest in the modes of CO$_2$. An extensive e-mail
exchange with Ad Huijser on the subject is gratefully acknowledged.
Wim Witteman provided an
analysis of the spectrum of CO$_2$ \cite{witteman}.
\appendix
\section{Generalized 1-dimensional  Propagator}
In this section we discuss the flow on the basis of the
propagator
\begin{equation} \label{A1}
  G(\rho-\rho') = b \exp(-a|\rho'-\rho|)
\end{equation}
The purpose of this propagator is to provide a soluble
approximation for the angular averaged equations. (The
1-dimensional version of the model is characterised by the
values $a=b=1$.)

For the consistency of the flow and the absorption equation we
used $E_3(0)=1/2$. So if $G(\rho-\rho')$ is emploid as
an approxmation to $E_3(\rho-\rho')$ we must set $b=1/2$.

The flow equation of this propagator involves the three flows
\begin{equation} \label{A2}
  I_0 (\rho) + I_\uparrow (\rho) - I_\downarrow (\rho) = F_{\rm out},
\end{equation}
with the expressions
\begin{equation} \label{A3}
  \left\{
    \begin{array}{rcl}
      I_0 (\rho) & = & \displaystyle I_0 \,r \, G(q-\rho) +
                       I_0 \frac{1-r}{2}, \\*[4mm]
      I_\uparrow (\rho) & = & \displaystyle r \int^q_\rho d\rho'
         G(\rho-\rho') j(\rho') + \frac{1-r}{2} \int^q_\rho d \rho'
                         j(\rho'), \\*[4mm]
      I_\downarrow (\rho) & = & \displaystyle r \int^\rho_0 d\rho'
         G(\rho-\rho') j(\rho') + \frac{1-r}{2} \int^\rho_0 d \rho'
                         j(\rho'). \\*[4mm]
    \end{array}
  \right.
\end{equation}

The flow equation is an equation for the absorption $j(\rho)$. The
absorption equation follows by differentiation of the flow equation
with respect to $\rho$. It has the form
\begin{equation} \label{A4}
  j(\rho)= ar \left(I_0 G(q - \rho) + \int^q_\rho d\rho'
    G(\rho-\rho') j(\rho') + \int^\rho_0 d\rho'
    G(\rho-\rho') j(\rho') \right).
\end{equation}
By applying the operator $\cal O$
\begin{equation} \label{A5}
  {\cal O} = \exp(-a \rho) \frac{d}{d \rho} \exp(2 a \rho)
  \frac{d }{d \rho} \exp(-a \rho)
\end{equation}
to both sides of the equation it obtains the form
\begin{equation} \label{A6}
  \frac{ d^2 j(\rho)}{d \rho^2} - a^2 j(\rho) = - a^2 r j(\rho).
\end{equation} 
 The solution of this absorption equation is a sum of
 two exponentials
 \begin{equation} \label{A7}
   j(\rho) = [a_+ \exp(a\alpha \rho) + a_- \exp(-a\alpha \rho)] F_{\rm out}
 \end{equation}
 with (as in Eq.~(\ref{e2}))
 \begin{equation} \label{A8}
   \alpha = \sqrt{1 -r}.
 \end{equation}
 The constants $a_\pm$ are found by
 inserting the expression (\ref{A7}) into the absorption equation
 (\ref{A5}). The result is a sum over four exponentials. Those involving
 $\exp(\pm \alpha \rho)$ vanish due to the expression $(\ref{A8})$
 for $\alpha$. The terms with $\exp(-\rho)$ give the condition
\begin{equation} \label{A9}
  a_+ (1 - \alpha) + a_- (1 + \alpha) = 0.
\end{equation} 
The terms with $\exp(\rho-q)$ give the relation
\begin{equation} \label{A10}
\frac{I_0}{F_{\rm out}} = a_+ \frac{\exp(a \alpha q)}{a(1-\alpha)} + 
a_- \frac{\exp(-a \alpha q)}{a(1+\alpha)}.
\end{equation}
One gets a third relation between $I_0$ and $F_{\rm out}$ by plugging
the solution Eq.~(\ref{A7}) into the flow equation.
This leads again to the two equations (\ref{A9}) and (\ref{A10})
and in addition the constant terms give the condition
\begin{equation} \label{A11} 
  \alpha^2\frac{I_0}{F_{\rm out}} + a_+\frac{\alpha}{2} [\exp(a \alpha q) +1] -
  a_-\frac{\alpha}{2} [\exp(-a \alpha q) +1] = 1.
\end{equation}
Using relation (\ref{A10}) in (\ref{A11}) one has with condition
(\ref{A9}) two equations for the two coefficients $a_\pm$. The
solution reads
\begin{equation} \label{A12}
  a_+ = \frac{2a(1+\alpha)}{\alpha \det(q,\alpha)},
  \quad\quad
  a_- = -\frac{2a(1-\alpha)}{\alpha \det(q,\alpha)}.
\end{equation}
with the expression for $\det(q,\alpha)$
\begin{equation} \label{A13}
  \det(q,\alpha) = \frac{1+\alpha}{1-\alpha} \exp(a q \alpha) +
   \frac{1-\alpha}{1+\alpha} \exp(-a q \alpha) +2 .
\end{equation} 
With the $a_\pm$ known as function of $\alpha$ (or $r$) and $q$
all the currents are known. A fast route to these expressions is
to substitute $I_0$ from Eq.~(\ref{A10}) into Eq.~(\ref{A11}) and
then solve with Eq.~(\ref{A9}) for $a_\pm$.

Using these formulae we find for the
downstream at the surface of the earth
\begin{equation} \label{A14}
  \frac{I_\downarrow (0)}{F_{\rm out}} =
  \left( \frac{1+\alpha}{\alpha} \exp (a \alpha q)
    - \frac{1-\alpha}{\alpha} \exp (-a \alpha q) -2\right)
  /\det(q,\alpha)
\end{equation}
The greenhouse factor $R$  equals
\begin{equation} \label{A15}
R=[1+2 I_\downarrow (0)/F_{\rm out}]^{1/4}
\end{equation}


\begin{thebibliography}{99}
\bibitem{thickness}
  As long as $q(\nu)$ is still a function of $\nu$
  the more common optical thickness $\rho = q \tau$ cannot be used
  as coordinate for the height.
\bibitem{lorentz} Far out a Lorentzian line goes
  over in a Doppler broadened line, but that region is less important
  since the absorption is there already very weak.
\bibitem{kirchhoff} The relation is often refered to as the
  law of Kirchhoff stating that the emissivity equals the
  absorptance. Here we use it only in a global way as conservation
  of energy.
\end{thebibliography}
\end{document}